\documentclass[twocolumn,preprintnumbers,amsmath,amssymb]{revtex4}

\usepackage{graphicx}
\usepackage{epstopdf}
\usepackage{dcolumn}
\usepackage{bm}

\begin{document}
\def\P{\mathbf{P}}
\def\Q{\mathbf{Q}}


\title{Towards a mathematical theory of meaningful communication}

\author{Bernat Corominas-Murtra$^{1,2,3}$, Jordi Fortuny$^{4}$, Ricard V. Sol\'e$^{2,3,5}$}

\affiliation{
$^1$ Section for Science of Complex Systems, Medical University of Vienna,  Vienna, Austria\\
$^2$ ICREA-Complex Systems Lab, Universitat Pompeu Fabra,  Barcelona, Spain\\
$^3$Institut de Biologia  Evolutiva. CSIC-UPF. Barcelona, Spain\\
$^4$Departament de Ling\"u\'istica  General. Universitat de Barcelona.  Barcelona, Spain\\
$^5$Santa Fe Institute,   Santa Fe, New Mexico USA
}

\begin{abstract}
Despite its obvious relevance, meaning has been outside most theoretical approaches to information in biology. 
As a consequence, functional responses based on an appropriate interpretation of signals has been 
replaced by a probabilistic description of correlations between emitted and received symbols. 
This assumption leads to potential paradoxes, such as the presence of a maximum information 
associated to a channel that would actually create completely wrong interpretations of the signals. 
Game-theoretic models of language evolution use this view of Shannon's theory, but other approaches 
considering embodied communicating agents show that the correct (meaningful) match resulting from 
agent-agent exchanges is always achieved and natural systems obviously solve the problem correctly. How can 
Shannon's theory be expanded in such a way that meaning -at least, in its minimal referential form- is properly incorporated? 
Inspired by the concept of {\em duality of the communicative sign} stated by the swiss linguist 
Ferdinand de Saussure, here we present a complete description of the minimal 
system necessary to measure the amount of information that is consistently decoded. 
Several consequences of our developments are investigated, such the uselessness of 
an amount of information properly transmitted for communication among autonomous agents.
\end{abstract}

\maketitle

\section{Introduction}

Major innovations in evolution have been associated with novelties in the ways 
information is coded, modified and stored by biological structures on multiple scales \cite{Schuster:2001}. 
Some of the major transitions involved the emergence of complex forms of communication, 
being human language the most prominent and difficult to explain \cite{Szathmary:1997}.  
The importance of information in biology has been implicitly recognised since the early developments of 
molecular biology, which took place simultaneously with the rise of computer science and information 
theory. Not surprisingly, many key concepts such as coding, decoding, transcription or translation were 
soon incorporated as part of the lexicon of molecular biology \cite{MaynardSmith:2000}.  

Communication among individual cells promoted multicellularity, which required the invention and 
diversification of molecular signals and their potential interpretations. Beyond genetics, 
novel forms of non-genetic information propagation emerged. In a later stage, the rise of 
neural systems opened a novel scenario to share communication with full richness \cite{Szathmary:1997}. 
Human language stands as the most complex 
communication system and, since communication deals with creation, 
reception and processing of information, understanding communication in 
information theoretic terms has become a major thread in our approach to the evolution of language. 
\begin{figure}
\includegraphics[width= 8 cm]{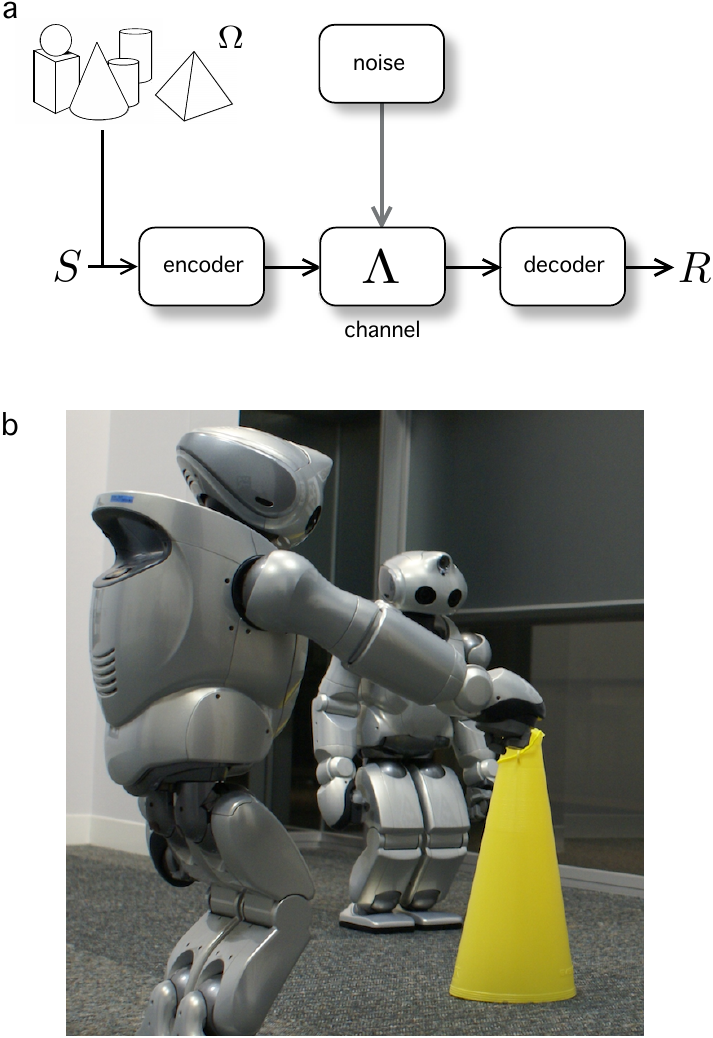}
\caption{In standard theory of information, as defined in Shannon's theory, 
a communication system (a) is described in terms of a sequential chain of steps 
connecting a source of messages (S) and a final receiver (R). The source can be 
considered linked to some external repertoire of objects ($\Omega$). An encoder and a decoder 
participate in the process and are tied through a channel $\Lambda$, subject to noise. 
The acquisition and evolution of a language, such as those emerging in artificial systems of interacting agents, such as robots (b) 
involves some additional aspects that are usually ignored in the original formulation of Shannon's 
approach. Those include the embodiment of agents and the necessary consistency emerging from the shared 
perception of the external world. }
\label{Fig:MinimalSyst}
\end{figure}

In its classical form, information theory (IT) has been formulated as a way of defining how 
signals sent and received through a given channel with no attention to their meaning. However, 
in all kinds of living systems, from cells sharing information about their external medium, individuals of a given species surviving 
in a world full of predators or when two humans or apes exchange signals, a crucial component 
beyond information is its meaningful content \cite{Hopfield:1994}. The distinction is very important, since information has been 
treated by theoreticians since Shannon's seminal work \cite{Shannon:1948} as a class of statistical object 
that measures correlations among sets of symbols, whereas meaning is inevitably tied to 
some sort of functional response with consequences for the fitness of the communicating agents. 
This standard scheme describing information transmission through a noisy channel \cite{Shannon:1948} is summarized in 
figure (1)a. The most familiar scenario would be described by a speaker (S) and a listener or receiver (R) 
having a conversation in a living room. The air carries the voice of the first and is the channel, which would 
be reliable (low or zero noise) if nothing except R and S were present. Instead, the channel will become more and more 
unreliable (noisy) as different sources of perturbation interfere. These can be very diverse, from 
air turbulence and children laughing to another conversation among different people. Consistently with any 
standard engineering design, Shannon's picture allows us to define efficient communication in terms somewhat similar to those used -for example- within electric transmission networks. In this case, a goal of the system design is minimizing 
the heat loss during the transmission process. Information is a (physically) less obvious quantity, but the approach 
taken by standard IT is quite the same. 

As a consequence of its statistical formulation, IT does not take into account 
"meaning" or "purpose" which, as noted by Peter Schuster \cite{Schuster:2001} are also difficult notions 
for evolutionary biology. Despite this limitation, it has been shown to successfully 
work in the analysis of correlations in biology \cite{Bialek:2012}. However, one 
undesirable consequence of this approach is that some paradoxical situations can emerge 
that contradict our practical intuition. An example is that a given pair of signals $s_1, s_2$ associated to two 
given objects or events from the external world can be "interpreted" by the receiver of the messages in a completely 
wrong way -"fire" and "water" for example to be understood, with all its consequences, as "water" and "fire", respectively. 
Measured from standard IT -see below- the information exchanged is optimal -even perfect- if "fire" ("water") is always interpreted 
as "water" ("fire"). In other words, full miscommunication can also score high as "efficient" within Shannon's approach. 
Therefore, the communicative sign as a {\em dual} entity that must be preserved as a whole in the communicative exchange. This crucial {\em duality sign} in communicative exchanges was already pointed out -with some conceptual differences to the version we will expose below- before the birth of information theory by the swiss linguist Ferdinand de Saussure in his acclaimed {\em Cours de Linguistique gen\'erale} \cite{Saussure:1916}. 

It seems obvious that meaning -and the connection to this to some signal, to create the dual entity- plays an essential role and has been shaped through evolution: 
"the message, the machinery processing the message and the context in which the message is evaluated are generated 
simultaneously in a process of coevolution" \cite{Schuster:2001}. 
In our bodies, proper recognition of invaders is essential to survival, and failures to recognizing the self 
and the non-self are at the core of many immune diseases \cite{Atlan:1987, Atlan:1998}. Similarly, learning processes 
associated to proper identification of predators and how to differentiate them from inmates are 
tied to meaningful information. Beyond the specific details associated to each system, correct information storing and sharing, and the relevance of meaning is well illustrated by its impact  on evolutionary dynamics. 
As pointed out in \cite{MaynardSmith:2000} we can say that, 
in biology, the coder is natural selection. 
In this way, the use of evolutionary game theoretic arguments has played a very important role in shaping evolutionary theory 
\cite{Hurford:1989, Nowak:1999, Nowak:2000, Plotkin:2000, Komarova:2004, Niyogi:2006}, 
but require some extension in order to properly account for meaningful information. 
Moreover, evolutionary robotics and the artificial evolution of protolanguages and proto-grammars is a unique 
scenario where such a framework naturally fits \cite{Cangelosi:2002, Steels:2001, Steels:2003, Steels:2003b, Steels:2005, Floreano:2007, Nolfi:2010}. Evolving robots capable of developing simple communication skills are able of acquiring a repertoire of appropriate signals, share them and interpret correctly the signals sent by other agents. The coherent development of a shared set of symbols that is correctly used -and thus where "meaning" is preserved. 
Such coherence results from the combination of a shared repertoire of signals together with a shared perception of 
the external world, as detected and perceived by the same class of sensing devices. 

In this paper we develop and describe an information-theoretic minimal system in which the signal is linked  to a referential value. In a nutshell, we are going to derive an information-theoretic measure able to grasp for the consistency of the shared information between agents, when the {\em meaning} is introduced as a primitive referential value attributed to one or more signals.

\section{{\em Meaning} in a minimal system}

We start this section describing the minimal system incorporating referential values for the sent signals -section \ref{Sec:Description}. Within this system, we show what is meant when we say that information theory is blind to any meaning of the message -section \ref{sec:ITMS}. Then, derive the amount of consistently decoded information among two given agents exchanging information of their shared world, thereby fixing the problem pointed out above -section \ref{Sec:DerivationConsistentInfo}- and derive some of its most salient properties, including the complete description of the binary symmetric channel within this new framework -section \ref{Sec:Properties}.

\subsection{The minimal system encompassing Referentiality}

\label{Sec:MinimalSyst}

Our minimal system to study the consistency of a given information exchange will involve two {\em autonomous communicative agents}, $\mathbf{A},\mathbf{B}$, a {\em channel}, $\Lambda$, and a {\em shared world}, $\Omega$. Agents exchange information about their shared world through the channel -see figure (\ref{Fig:MinimalSyst}). Now we proceed to describe it in detail.

\subsubsection{Description}
\label{Sec:Description}
An {\em agent}, $\mathbf{A}$, is defined as a pair of computing devices,
\[
\mathbf{A}\equiv \{\P^\mathbf{A},\Q^\mathbf{A}\},
\]
where $\P^\mathbf{A}$ is the coder module and $\Q^\mathbf{A}$ is the decoder module. The
shared world is defined by  a random variable $X_{\Omega}$ which takes
values on the set of events, $\Omega$,
$$  \Omega=\{m_1,...,m_n\}$$
being  the  (always  non-zero)  probability associated  to  any  event
$m_k\in\Omega$ defined  by $p(m_k)$.   The coder module,  $\P^{\mathbf{A}}$, is
described by a mapping from $\Omega$ to the set 
$$ {\cal S}=\{s_1,...,s_n\}$$
to be identified as the set of signals. To start with, here we assume
$|\Omega|=|{\cal  S}|=n$, unless the contrary is indicated. The coder module 
is represented by  a mapping defined according  to a matrix of conditional probabilities $\P^{\mathbf{A}}$, having elements
 $\P^{\mathbf{A}}_{ij}=\mathbb{P}^{\mathbf{A}}(s_j|m_i)$,
and satisfying the normalization conditions, namely,
for all $m_i\in \Omega$, $\sum_{j\leq n}\P^{\mathbf{A}}_{ij}=1$. The outcome of the coding process is depicted by the random variable $X_s$, taking values over ${\cal S}$ following a probability distribution 
\[
q(s_i)=\sum_{j\leq n} p(m_j) \P^{\mathbf{A}}_{ji}.
\]
The channel $\Lambda$ is characterized  by the $n\times n$ matrix of
conditional  probabilities $\Lambda$,
with matrix elements $\Lambda_{ij}=\mathbb{P}_{\Lambda}(s_j|s_i)$. 
The random variable $X'_s$ describes the output of the  composite system world+coder$+$channel,  thereby
taking values on the set
${\cal S}$, and follows the probability distribution $q'$, defined as
\[
q'(s_i)=\sum_k p(m_k)\sum_{j\leq n}\P^{\mathbf{A}}_{kj}\Lambda_{ji}.
\]
Finally, the decoder  module is a computational device  described by a
mapping from ${\cal  S}$ to ${\Omega}$; i.e it  receives ${\cal S}$ as
the  input set,  emitted by  another  agent through  the channel,  and
yields as output  elements of the set $\Omega$.   $\Q^{\mathbf{A}}$ is completely
defined by its transition probabilities, namely,
$\Q^{\mathbf{A}}_{ik}=\mathbb{P}^{\mathbf{A}}(m_k|s_i)$,
which satisfies the normalization conditions, i.e.,
for all $s_i\in {\cal S}$, $\sum_{k\leq n}\Q^{\mathbf{A}}_{ik}=1$.
We emphasize  the assumption that,
in a given agent  ${\mathbf{A}}$, following \cite{Nowak:1999, Plotkin:2000} but
not   \cite{Hurford:1989,  Komarova:2004}   there  is   a   priori  no
correlation between $\P^{\mathbf{A}}$ and $\Q^{\mathbf{A}}$.  

Now suppose that we want to study the information transfer between two agents sharing the world. Let us consider $\mathbf{A}$ the decoder agent and $\mathbf{B}$ the decoder one, although we emphasize that both agents can perform both tasks. Agent ${\mathbf{B}}$ tries to reconstruct $X_{\Omega}$ from the information received from $\mathbf{A}$. The description of $\Omega$ made by agent $\mathbf{B}$ is depicted by the random
variable  $X'_{\Omega}$,  taking  values   on  the  set  $\Omega$  and
following the probability distribution $p'$, which takes the form:
\begin{equation}
p'(m_i)\equiv \sum_{l\leq n}p(m_l)\mathbb{P}_{{\mathbf{A}}{\mathbf{B}}}(m_i|m_l),
\label{eq:p'}
\end{equation}
where 
\[
\mathbb{P}_{ {\mathbf{A}}{\mathbf{B}}}(m_i|m_l)=\sum_{j,r\leq n}\P^{\mathbf{A}}_{lj}\Lambda_{jr}\Q^{\mathbf{B}}_{ri}.
\]
From which we can naturally derive the joint probabilities, $\mathbb{P}_{{\mathbf{A}\mathbf{B}}}(m_i, m_j)$ as follows:
\begin{eqnarray}
\mathbb{P}_{{\mathbf{A}} {\mathbf{B}}}(m_i,m_j)&=&\sum_{l,r}p(m_j)\P^{\mathbf{A}}_{jl}\Lambda_{lr}\Q^{\mathbf{B}}_{ri}.
\label{joint}
\end{eqnarray}
We say that $X'_{\Omega}$ is the {\em reconstruction} of the shared world, $X_{\Omega}$, made by agent $\mathbf{B}$ from the collection of messages sent by $\mathbf{A}$. Summarizing, we thus have a composite system where the behavior at every step is described by a random variable, from the description of the world, $X_{\Omega}$ to its reconstruction, $X'_{\Omega}$ -see figure (\ref{Fig:MinimalSyst}a):
\[
\Omega \overbrace{\longrightarrow}^{X_{\Omega}\sim p}\mathbf{A}\overbrace{\longrightarrow}^{X_s\sim q}\Lambda\overbrace{\longrightarrow}^{X'_s\sim q'}\mathbf{B}\overbrace{\longrightarrow}^{X'_{\Omega}\sim p'}\Omega.
\]

At this point, it is convenient to introduce, for the sake of simplicity, some new notation. We will define two matrices, namely $\mathbf{J}({\mathbf{A}\mathbf{B}})$ and $\Lambda({\mathbf{A}\mathbf{B}})$ in such a way that $J_{ij}({\mathbf{A}\mathbf{B}})\equiv \mathbb{P}_{\mathbf{A}\mathbf{B}}(m_i,m_j)$ and $\Lambda_{ij}({\mathbf{A}\mathbf{B}})\equiv\mathbb{P}_{\mathbf{A}\mathbf{B}}(m_j|m_i)$. Finally, we will define the probability distribution $\Lambda_i({\mathbf{A}}{\mathbf{B}})\equiv \{\Lambda_{i1}({\mathbf{A}\mathbf{B}}),...,\Lambda_{in}({\mathbf{A}\mathbf{B}})\}$. This new notation will enable us to manage formulas in a compact way.

\subsubsection{Information-theoretic aspects of this minimal system}
\label{sec:ITMS}
Now we explore the behavior of mutual information in this system. Detailed definitions of information-theoretic functionals used in this subsection are provided in appendix A.
Under the above described framework we have two relevant  random variables, namely {\em the world} $X_{\Omega}$ and the {\em reconstruction} of the world $X'_{\Omega}$. Its  mutual information  $I(X_{\Omega}:X'_{\Omega})$ is
defined as \cite{Shannon:1948, Ash:1990, Thomas:2001}:
\begin{equation}
I(X_{\Omega}:X'_{\Omega})=H(X_{\Omega})-H(X_{\Omega}|X'_{\Omega}).
\label{InfoShan}
\end{equation}
The above expression has an equivalent formulation, namely
\begin{eqnarray}
I(X_{\Omega}:X'_{\Omega})=\sum_{i,j\leq n}J_{ij}(\mathbf{AB})\log \frac{J_{ij}(\mathbf{AB})}{p(m_i)q(m_j)},
\label{InfoKL}
\end{eqnarray}
where the right side of the above equation can be identified as the {\em Kullback-Leibler} divergence between distributions $J(\mathbf{AB})$ and $p\cdot q$:
\begin{equation}
I(X_{\Omega}:X'_{\Omega})=D(J(\mathbf{AB})||p\cdot q).
\end{equation} 
Within this formulation, the mutual information is the amount of accessory bits needed to describe the composite system $X_{\Omega},X'_{\Omega}$ taking as the reference the distribution $p\cdot q$, which supposes no correlation between $X_{\Omega}$ and $X'_{\Omega}$.  

Let us emphasize a feature of mutual information which is relevant for our purposes. As is well-known, $\max I(X_{\Omega},X'_{\Omega})\leq H(X_{\Omega})$, and equality holds if there is no ambiguity in the information processing process,  meaning that the process is {\em reversible}, in logical terms. Thus, every event $m_i\in\Omega$ has to be decoded with probability $1$ to some event $m_j\in\Omega$ which, in turn, must not be the result of the coding/decoding process of any other event.
In mathematical terms, this means that $\P^{\mathbf{A}},\Q^{\mathbf{B}}, \Lambda \in \Pi_{n\times n}$,
being $\Pi_{n\times n}$ the set of $n\times n$ permutation matrices, which are the matrices in which every file and column contains $n-1$ elements equal to $0$ and one element equal to $1$ -see appendix B.  It is worth emphasizing that $\delta_{n\times n}$, the $n\times n$ identity matrix is itself a permutation matrix. Notice that if $\Lambda ({\mathbf{A}\mathbf{B}})\neq\delta$ some symbol $m_i$ sent by the source is decoded as a different element $m_j$. This shift has no impact on the information measure $I(X_{\Omega}:X'_{\Omega})$ and this is one of the reasons by which it is claimed that {\em the content of the message} is not taken into account in the standard information measure. Actually, it is straightforward to show -see Appendix B- that only $n!$ out of the $(n!)^3$ configurations leading to the maximum mutual information also lead to a fully consistent reconstruction -a reconstruction where referential value is conserved. This mathematically shows that, for autonomous agents exchanging messages, mutual information is a weak indicator of communicative success.
\begin{figure}
\includegraphics[width= 8 cm]{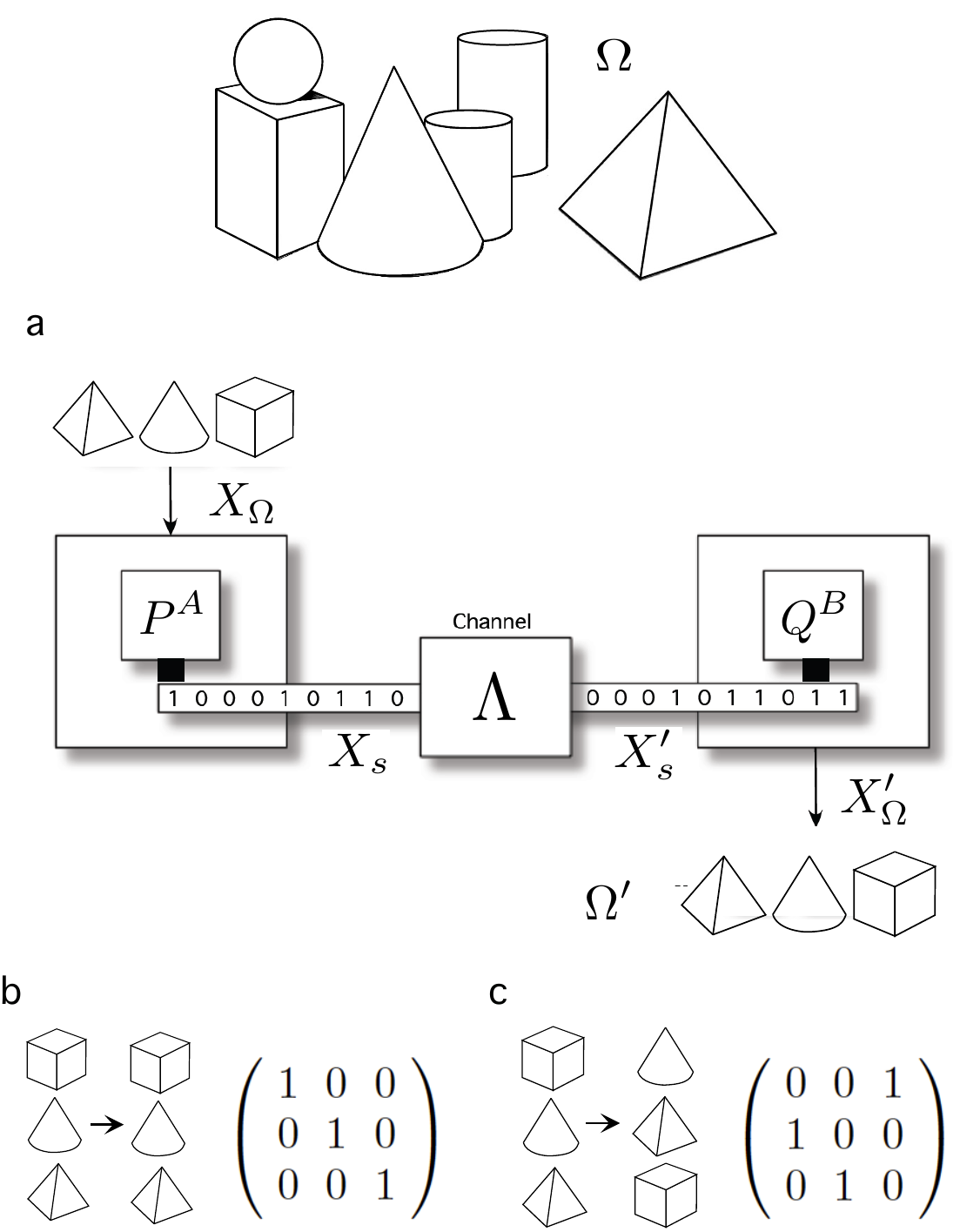}
\caption{The minimal communicative system to study the conservation of
  referentiality (a): A shared world,  whose events are the members of
  the  set $\Omega$  and  whose  behavior is  governed  by the  random
  variable $X_{\Omega}$.   A coding  engine, $\P^{\mathbf{A}}$, which  performs a
  mapping between ${\Omega}$ and the  set of signals ${\cal S}$, being
  $X_s$  the random  variable describing  the behavior  of the  set of
  signals obtained after coding.  The channel, $\Lambda$, may be noisy
  and, thus,  the input  of the decoding  device, $\Q^\mathbf{B}$,  depicted by
  $X'_s$, might  be different from  $X_s$.  $\Q^\mathbf{B}$ performs  a mapping
  among  ${\cal  S}$  and   $\Omega$  whose  output  is  described  by
  $X'_{\Omega}$.  Whereas the mutual information provides us a measure
  of  the  relevance  of   the  correlations  among  $X_{\Omega}$  and
  $X'_{\Omega}$,  the  {\em   consistent  information}  evaluates  the
  relevance  of the information  provided by  consistent pairs  on the
  overall  amount  of  information.   In  this context,  from  a  pure
  information-theoretical  point of  view, situations  like b)  and c)
  could  be indistinguishable.  By defining  the  so-called consistent
  information we  can properly differentiate  b) and c)  by evaluating
  the degree of consistency of input/output pairs -see text.}
\label{Fig:MinimalSyst}
\end{figure}

\subsection{Derivation of {\em consistent information}}

\label{Sec:DerivationConsistentInfo}
Now we have a complete description of the minimal system able to encompass referential values for the sent signals. It is the objective of this section to derive an information-theoretic measure, different from mutual information, able to evaluate the amount of consistently decoded information.

\subsubsection{Preliminaries}
The rawest evaluation of the amount of consistently decoded pairs is found by averaging the probability of having a consistent coding/decoding process during an information exchange between agent $\mathbf{A}$ and agent $\mathbf{B}$. This corresponds to the view of an external observer simply counting events only taking into account wether they are consistently decoded or not. This probability, to be named $\theta_{\mathbf{AB}}$, is obtained by summing the probability of having consistent input output pair, i.e.:
\begin{equation}
\theta_{\mathbf{AB}}={\rm tr}J(\mathbf{AB})=\sum_{i\leq n}J_{ii}(\mathbf{AB}).
\label{Eq:Theta}
\end{equation}
This formula has been widely used as a communicative payoff for an evolutionary dynamics in which consistent communication has a selective advantage. We observe that the probability of error $p_e(\mathbf{AB})$ in this scenario is just $p_e(\mathbf{AB})=1-\theta_{\mathbf{AB}}$. Therefore, thanks to Fano's inequality -see Appendix A-, we can relate this parameter to the information-theoretic functionals involved in the description of this problem, namely:
\begin{equation}
\theta_{\mathbf{AB}}\leq 1-\frac{H(X'_{\Omega}|X_{\Omega})}{\log(n-1)}.
\label{eq:Fano}
\end{equation}
From this parameter, we can build another, a bit more elaborated functional. We are still under the viewpoint of the external observer who is now interested in the fraction of information needed to describe the composite system $X_{\Omega},X'_{\Omega}$ that comes from consistent input/output pairs when information is sent from $\mathbf{A}$ to $\mathbf{B}$. This fraction, to be named $\sigma_{\mathbf{AB}}$ is:
\[
\sigma_{\mathbf{AB}}=\frac{{\rm tr}(J(\mathbf{AB})\log J(\mathbf{AB}))}{H(X_{\Omega},X'_{\Omega})}.
\]
We observe that the above quantity is symmetrical in relation to $X_{\Omega}$ and $X'_{\Omega}$. These two estimators provide global indicators of consistency of the informative exchange.

\subsubsection{Consistent Information}

However, we can go further and ask us {\em how much of the information from the environment is consistently decoded by agent $\mathbf{B}$ when receiving data from $\mathbf{A}$}. To start with, we observe that, since $J_{ij}(\mathbf{AB})=p(m_i)\Lambda_{ij}(\mathbf{AB})$, we can rewrite equation (\ref{InfoKL}) as:
\begin{eqnarray}
I(X_{\Omega},X'_{\Omega})&=&\sum_{i\leq n} p(m_i)\sum_{j\leq n}\Lambda_{ij}(\mathbf{AB})\log\frac{\Lambda_{ij}(\mathbf{AB})}{p'(m_j)}\nonumber\\
&=&\sum_{i\leq n}p(m_i) D(\Lambda_i (\mathbf{AB})||p').\nonumber
\end{eqnarray}
Knowing that $D(\Lambda_i(\mathbf{AB})||q)$ is the {\em information gain} associated to element $m_i$, $p(m_i)D(\Lambda_i(\mathbf{AB})||q)$ is its weighted contribution to the overall information measure. If we are interested on the amount of this information that is consistently referentiated, we have to add an ``extra" weight to $p(m_i)$, namely $\Lambda_{ii}(\mathbf{AB})$, which is just the probability of having $m_i$ both at the input of the coding process and at the output. Thus, since
\[
\Lambda_{ii}(\mathbf{AB})p(m_i) D(\Lambda_i(\mathbf{AB})||q)=J_{ii}(\mathbf{AB})D(\Lambda_i(\mathbf{AB})||p'),
\]
the amount of {\em consistent information} conveyed from agent ${\mathbf{A}}$ to agent ${\mathbf{B}}$, ${\cal I}({\mathbf{AB}})$, will be:
\begin{equation}
{\cal I}({\mathbf{AB}})=\sum_{i\leq n}J_{ii}(\mathbf{AB}) D(\Lambda_i(\mathbf{AB})||p').
\label{eq:ConsistentKullback}
\end{equation}
Since this is the most important equation of the text, we rewrite it using standard probability notation:
\begin{widetext}
\begin{equation}
{\cal I}({\mathbf{AB}})=\sum_{i,j\leq n}\mathbb{P}_{\mathbf{A}\mathbf{B}}(m_i,m_i) 
\sum_{j\leq n}\mathbb{P}_{\mathbf{AB}}(m_j|m_i)\log \left ( \frac{\mathbb{P}_{\mathbf{AB}}(m_j|m_i)}{p'(j)} \right ).
\label{eq:Explicit}
\end{equation}
\end{widetext}
We observe that the dissipation of consistent information is due to both standard noise $H(X_{\Omega}|X'_{\Omega})$, and another term, which is subtracted to $I(X_{\Omega}:X'_{\Omega})$, accounting for the loss of referentiality. Using equations (\ref{InfoShan}, \ref{InfoKL}) and (\ref{eq:ConsistentKullback}) we can isolate this new source of information dissipation, the {\em referential noise}, $\nu(\mathbf{AB})$, leading to:
\begin{equation}
\nu(\mathbf{AB})=\sum_{i\leq n}D(\Lambda_i(\mathbf{AB})||q)\left[\sum_{k\neq i}J_{ik}(\mathbf{AB})\right].\nonumber
\end{equation}
Therefore, the total loss of referential information or {\em total noise} will be described as
\[
\eta(\mathbf{AB})\equiv H(X_{\Omega}|X'_{\Omega})+\nu(\mathbf{AB}).
\]
The above expression enables us to rewrite equation (\ref{eq:ConsistentKullback}) as:
\begin{equation}
{\cal I}({\mathbf{AB}})=H(X_{\Omega})-\eta(\mathbf{AB}),
\label{Eq.ConsShannonForm}
\end{equation}
which mimics the classical Shannon Information, now with a more restrictive noise term. Interestingly, the above expression is not symmetrical: the presented formalism differentiates between the world $X_{\Omega}$ and its reconstruction, $X'_{\Omega}$.
If we take into account that, attending the definition we provided for an autonomous communicating agent, the information can flow both $\mathbf{A}\rightarrow\mathbf{B}$ and $\mathbf{B}\rightarrow\mathbf{A}$, we can compute the average success of the communicative exchange between $\mathbf{A}$ and $\mathbf{B}$, ${\cal I}(\mathbf{A:B})$, as:
\begin{equation}
{\cal I}(\mathbf{A:B})=H(X_{\Omega})-\frac{1}{2}(\eta(\mathbf{AB})+\eta(\mathbf{BA})).
\label{eq:Symmetrized}
\end{equation}
${\cal I}(\mathbf{A}:\mathbf{B})$ is the {\em consistent information about the $\Omega$ shared by agents $\mathbf{A}$ and $\mathbf{B}$}. We observe that, now, the above expression is symmetrical, ${\cal I}(\mathbf{A:B})={\cal I}(\mathbf{B:A})$, because both agents share the same world, represented by $X_{\Omega}$. We remark that the above expression is an information-theoretic functional {\em between two communicating agents}, it is not an information-measure among two random variables, as mutual information is. This equation quantifies the communication success between two minimal communicating agents $\mathbf{A, B}$ transmitting messages about a shared world.

\subsection{Properties}
\label{Sec:Properties}
In this section we present several important consequences that can be derived from the study of the presented consistent information. The rigorous and complete proofs of them can be found in the appendix, as well as a brief discussion about the self-consistency of agents.

\subsubsection{The Binary Symmetric Channel}
\label{sec:BSC}
\begin{figure}
\includegraphics[width= 8 cm]{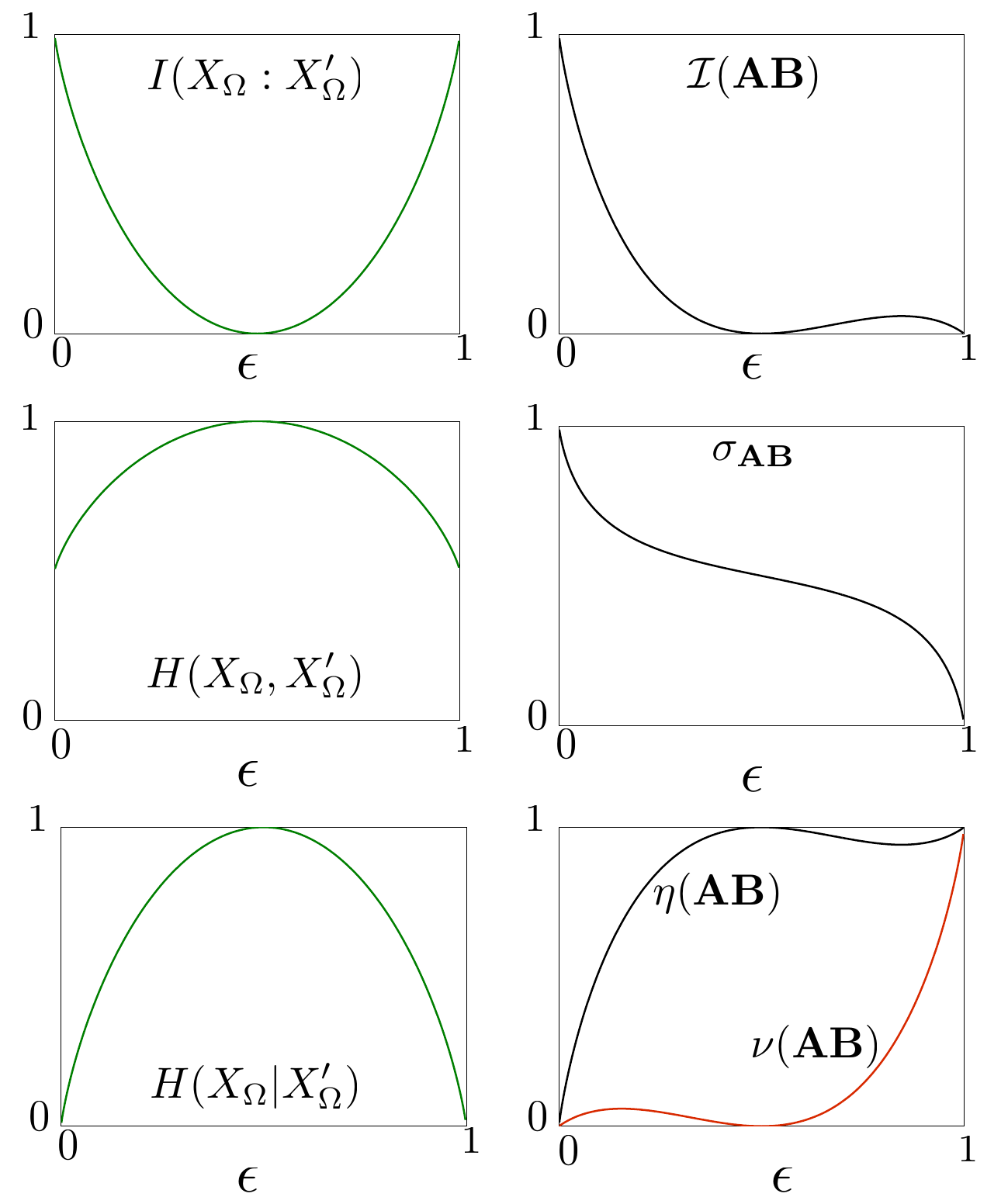}
\caption{The binary symmetric channel when we enrich the communication system with a referential set shared by coder and decoder agent. Plots correspond to the different values of the binary symmetric channel along  $\epsilon$, the referential shift parameter, from $\epsilon=0$ (total information with no loss of referentialty) to $\epsilon=1$ (total information with total loss of referentiality). At left, from top to bottom, we have the classical, well known plots of $I(X_{\Omega}:X'_{\Omega})$, $H(X_{\Omega},X'_{\Omega})$ (normalized to $1$) and $H(X_{\Omega}|X'_{\Omega})$. At right, we have the equivalent ones accounting for the referentiality conservation, namely, on top, ${\cal I}(\mathbf{A}\mathbf{B})$, next, $\sigma_{\mathbf{AB}}$ and in the last plot, we have $\eta(\mathbf{A}\mathbf{B})$ (black line) and $\nu(\mathbf{AB})$ (red line) -see section \ref{Sec:DerivationConsistentInfo}. units are given in bits. We observe that both $I(X_{\Omega}:X'_{\Omega})$ (and $H(X_{\Omega}:X'_{\Omega})$) have a symmetric behavior, with a minimum (maximum) at $\epsilon=\frac{1}{2}$ (total uncertainty). On the contrary, ${\cal I}(\mathbf{A}\mathbf{B})$ does not show a symmetric behavior, showing two minima, at $\epsilon=\frac{1}{2}$ and at $\epsilon=1$. There is  a local maxima at about $\epsilon\approx 0.85$ which is a by-product of the combination of the loss of uncertainty of the system and with a small but non-vanishable degree of referentiality conservation. }
\label{Fig:Informations}
\end{figure}

We first consider the simplest case, from which we can easily extract analytical conclusions that help us to gain intuition, the {\em Binary Symmetric Channel} with uniform input probabilities. We are concerned with a world $\Omega$ having two events such that $p(1)=p(2)=1/2$, two agents $\mathbf{A}$
 and $\mathbf{B}$ sharing information about this world, and a binary channel, $\Lambda$. The agents' and channel configuration are assumed to be of the following form:
\[
\Lambda(\mathbf{A}\mathbf{B})=\left(\begin{array}{cc}1-\epsilon & \epsilon\\ \epsilon & 1-\epsilon \end{array}\right),
\]
being $\Lambda(\mathbf{A}\mathbf{B})=\P^{\mathbf{A}}\Lambda\Q^{\mathbf{B}}$, as defined in section \ref{Sec:Description}. 
We will refer to $\epsilon$ as the {\em referential shift}, which is the probability that a given event is wrongly decoded in the reconstruction of $\Omega$. In this minimal system all functionals can be easily evaluated. First, we have that $I(X_{\Omega}, X'_{\Omega})=1-H(\epsilon)$, and that $\theta_{\mathbf{AB}}=1-\epsilon$, being $H(\epsilon)$ the entropy of a {\em Bernouilli process} having parameter $\epsilon$ -See appendix A. This leads to the following expression of the consistent information: 
\begin{equation}
{\cal I}(\mathbf{AB})=\theta_{\mathbf{AB}}(1-H(\epsilon))=\theta_{\mathbf{AB}}I(X_{\Omega}, X'_{\Omega}).
\label{eq:BinaryInfo}
\end{equation}
We can also easily compute $\sigma_{\mathbf{AB}}$:
\[
\sigma_{\mathbf{AB}}=\theta_{\mathbf{AB}}\frac{1-\log\theta_{\mathbf{AB}}}{2-H(\epsilon)}.
\]
The behavior of consistently decoded information is shown in figure (\ref{Fig:Informations}). In these plots we confronted the behavior of $I(X_{\Omega},X'_{\Omega})$, $H(X_{\Omega}, X'_{\Omega})$ and $H(X_{\Omega}, X'_{\Omega})$ with their analogous counterparts when referentiality is taken into account, nalemy ${\cal I}(\mathbf{AB})$ and $\sigma_{\mathbf{AB}}$ and $\nu({\mathbf{AB}})$ (and $\eta({\mathbf{AB}})$) respectively. We can observe the symmetric behavior of  the first ones against $\epsilon$, which highlights the total insensibility to referentiality conservation of these classical measures. instead, we observe that ${\cal I}(\mathbf{AB})$, $\sigma_{\mathbf{AB}}$, $\eta({\mathbf{AB}})$ and $\nu(\mathbf{AB})$ grasp the loss of referentiality conservation, showing a non-symmetric behavior with a generally decreasing trend as long as referentiality is lost.

\subsubsection{Decrease of information due to referential looses}

One interesting consequence of equation (\ref{eq:BinaryInfo}) is that, except very restricted situations, the presence of noise has a negative impact on the value of the consistent information, leading to the general conclusion that:
\begin{equation}
{\cal I}(\mathbf{AB})< I(X_{\Omega}:X'_{\Omega}).
\label{eq:IAB<I}
\end{equation}
This latter inequality shows that, in most cases, in the absence of designer, some of the information properly transmitted is actually useless for communication in a framework of autonomous agents.
As shown in the appendix, it is not difficult to show that the strict inequality holds in general. Indeed, the above relation becomes equality only in the very special case where there is perfect a matching between the two agents (i.e.: $\Lambda(\mathbf{A}\mathbf{B})=\delta_{n\times n}$, being $\delta_{n\times n}$ the $n\times n$ identity matrix.) or trivially, in the case where $I(X_{\Omega}:X'_{\Omega})=0$.

But we can go further. Let us consider that we know that the system displays a given value of  $I(X_{\Omega}:X'_{\Omega})$. We also know, by assumption, $H(X_{\Omega})$. We thus can easily derive $H(X_{\Omega}|X'_{\Omega})$ by simply computing $H(X_{\Omega})-I(X_{\Omega}:X'_{\Omega})$. Can we bound the value of ${\cal I}(\mathbf{AB})$ under such conditions? As happens with many problems of information theory, the general case is hard, even impossible to deal with. Instead, several approaches can be done assuming some special but illustrative cases. Let us assume the paradigmatic configuration in which $(\forall m_i\in \Omega) p(m_i)=1/n$ and  where $\Lambda(\mathbf{AB})$ acts as a symmetric channel. In this case, we have that ${\cal I}(\mathbf{AB})\leq\theta_{\mathbf{AB}}I(X_{\Omega}:X'_{\Omega})$, where 
\[
\theta_{\mathbf{AB}}\leq \frac{I(X_{\Omega}:X'_{\Omega})}{H(X_{\Omega})},
\]
and, therefore: 
\begin{eqnarray}
{\cal I}(\mathbf{AB})<\frac{I^2(X_{\Omega}:X'_{\Omega})}{H(X_{\Omega})}+1.
\label{eq:Inequality2}
\end{eqnarray}
(See Appendix for the details of the above derivations). This tells us, after some algebra, that in this framework,
\begin{eqnarray}
\eta(\mathbf{AB})\geq 2H(X_{\Omega}|X'_{\Omega})-\frac{H^2(X_{\Omega}|X'_{\Omega})}{H(X_{\Omega})}+1\nonumber.
\end{eqnarray}
Therefore, for $H(X_{\Omega})\gg H(X_{\Omega}|X'_{\Omega})$, we have that
$\eta(\mathbf{AB})\gtrsim 2H(X_{\Omega}|X'_{\Omega})$, leading to
\[
{\cal I}(\mathbf{AB})\lesssim H(X_{\Omega})-2H(X_{\Omega}|X'_{\Omega})
\]
and, for example, for the case in which $H(X_{\Omega})\approx 2H(X_{\Omega}|X'_{\Omega})$ we have that:
\begin{eqnarray}
{\cal I}(\mathbf{AB})&\lesssim & H(X_{\Omega})-\frac{3}{2}H(X_{\Omega}|X'_{\Omega})\nonumber\\
&=&\frac{1}{2}I(X_{\Omega}, X'_{\Omega}),\nonumber
\end{eqnarray}
The above examples enable us to illustrate the strong impact of noise on the conservation of  the referential value within a communication exchange -stronger than the one predicted by standard noise.

\section{Discussion}

Shannon's information theory had a great, almost immediate impact in all sorts 
of areas, from engineering and genetics to psychology or language studies \cite{Gleick:2012}. 
It also influenced the work of physicists, particularly those exploring the foundations 
of thermodynamics, who found that the entropy defined by Shannon provided powerful 
connections with statistical mechanics, particularly in terms of correlations. It is mainly at that 
level -i. e. the existence of correlations among different subsystems of a given system-
that the use of information theory has been shown to be useful. But correlations do not ensure 
a crucial type of coherence that seems necessary when dealing with meaningful communication: the 
preservation of referentiality.  

In this paper we have addressed a specially relevant problem, namely the development 
of an information-theoretic framework able to preserve meaning. This is a first step towards 
a more general goal, namely defining an evolutionary theory of language change 
including referentiality as an explicit component. We have shown that, if only the consistent 
information is considered, its value is significantly lower than mutual information in noisy scenarios.
We have derived the analytical form of a consistent information, which includes, along with the standard noise term, a referential noise. 
Our information measure defines a non-symmetrical function and properly weights 
the -more strict- requirement of consistency. We have illustrated our general results by means of the analysis of 
a classical, minimal scenario defined by the binary symmetric channel. The approach taken here should be 
considered as the formally appropriate framework to study the evolution of shared communication 
among embodied agents, where the presence of consistency is inevitable due to the 
shared perception constraints. Moreover, it might also be useful as a consistent mathematical 
framework to deal with cognitive-based models of brain-language evolution \cite{Deacon:1997, 
Christiansen:2003, Bickerton:1990}.

\vspace{0.5 cm}

{\bf Acknowledgments}

\vspace{0.25 cm}

We thank the members of the CSL for useful discussions. BC-M thanks 
Kepa Ruiz-Mirazo for helpful discussions on the topic. This work was supported 
by the Fundaci\'on Botin and by the Santa Fe Institute and the Austrian Fonds zur
F\"orderung der Wissenschaftlichen Forschung project "Quantifying socioeconomic 
multiplex networks in a massive multiplayer online game" KPP23378FW (to BC-M.).

\vspace{0.25 cm}

\appendix 

\section{Definitions}
\subsection{Information Theoretic Functionals}

The following definitions are intended to be minimal. We refer the interested reader to any standard textbook on information theory, such as \cite{Ash:1990} or \cite{Thomas:2001}.

.-Given a random variable $X_{\Omega}$ taking values over the set $\Omega$ following a probability distribution $p$,
\[
H(X_{\Omega})=-\sum_{i\leq n}p(m_i)\log p(m_i)
\]
is the standard {\em Shannon} or {\em statistical} entropy. 
 
.-Given two random variables, $X_{\Omega}$ and $X'_{\Omega}$, 
\[
 H(X_{\Omega}|X'_{\Omega}),=-\sum_{\i\leq n}q(m_i)\sum_{j\leq n}\mathbb{P}(m_j|m_i)\log \mathbb{P}(m_j|m_i)
\] 
is  the {\em conditional} entropy of $X_{\Omega}$ with respect $X'_{\Omega}$, being, in that case, $\mathbb{P}(m_j|m_i)\equiv \mathbb{P}(X_{\Omega}=m_j|X'_{\Omega}=m_i)$. Additionally, 
\[
H(X_{\Omega, X'_{\Omega}})-\sum_{\i\leq n}q(m_i)\sum_{j\leq n}\mathbb{P}(m_j,m_i)\log \mathbb{P}(m_j,m_i)
\]
where $\mathbb{P}(m_j,m_i)\equiv\mathbb{P}(X_{\Omega}=m_i, X'_{\Omega}=m_j)$ is the {\em joint entropy} of the two random variables $X_{\Omega}$, $X'_{\Omega}$.

.-Given two probability distributions $\pi_1,\pi_2$ defined over the set $\Omega$, the {\em Kullback-Leibler} divergence of {\em relative entropy} of $\pi_1$ with respect $\pi_2$ is:
\begin{equation}
D(\pi_1||\pi_2)=\sum_{i\leq n}\pi_1(x_i)\log\frac{\pi_1(x_i)}{\pi_2(x_i)},\nonumber
\end{equation}
which is the amount of extra information we need to describe $\pi_1$ taking as the reference  distribution $\pi_2$. 

.- {\em Fano}'s inequality. The probability of error in decoding is bounded satisfies the following inequality:
\[
p_e\geq \frac{H(X_{\Omega}|X'_{\Omega})-1}{\log(n-1)}.
\]

.- A Bernoulli process is a stochastic process described by a random variable $X$ taking value in the set $A=\{0,1\}$, being $p(0)=1-\epsilon$ and $p(1)=\epsilon$. $\epsilon$ is the {\em parameter} of the Bernoulli process. Its entropy $H(X)$ is commonly referred as $H(\epsilon)$, since it only depends on this parameter:
\[
H(\epsilon)=-(1-\epsilon)\log(1-\epsilon)-\epsilon\log\epsilon.
\]

\subsection{Permutation matrices}

A {\em permutation matrix } is
a square matrix which has exactly one entry equal to 1 in each row and
each  column and  0's elsewhere.   For example,  if $n=3$,  we  have 6
permutation matrices, namely:
\begin{equation}
\left(
\begin{array}{lll}
1&0&0\\
0&1&0\\
0&0&1
\end{array}
\right),
\left(
\begin{array}{lll}
1&0&0\\
0&0&1\\
0&1&0
\end{array}
\right),
\left(
\begin{array}{lll}
0&1&0\\
1&0&0\\
0&0&1
\end{array}
\right),\nonumber
\end{equation}
\begin{equation}
\left(
\begin{array}{lll}
0&0&1\\
1&0&0\\
0&1&0
\end{array}
\right),
\left(
\begin{array}{lll}
0&1&0\\
0&0&1\\
1&0&0
\end{array}
\right),
\left(
\begin{array}{lll}
0&0&1\\
0&1&0\\
1&0&0
\end{array}
\right).\nonumber
\end{equation}
The  set   of  $n\times  n$   permutation  matrices  is   indicated  as
$\Pi_{n\times    n}$    and    it    can    be    shown    that,    if
$\mathbf{A}\in\Pi_{n\times                                         n}$,
$\mathbf{A}^{-1}=\mathbf{A}^T\in\Pi_{n\times      n}$      and,     if
$\mathbf{A},\mathbf{B}\in\Pi_{n\times       n}$,      the      product
$\mathbf{A}\mathbf{B} \in  \Pi_{n\times n}$. Furthermore,  it is clear
that $\delta_{n\times n}\in \Pi_{n\times n}$, being $\delta$ the identity matrix or {\em Kronecker} symbol, defined as $\delta_{ij}=1$ if $i=j$ and $\delta_{ij}=0$, otherwise. 

\section{Inequalities}

We present the inequalities described in the main text in terms of three lemmas on the upper bounds of ${\cal I}(\mathbf{AB})$. The first one concerns inequality (\ref{eq:IAB<I}). The second one is general and supports the third, which proves inequality (\ref{eq:Inequality2}):
$\\$

{\em Lemma 1}.- Let $\mathbf{AB}$ be two agents sharing the world $\Omega$ as described in section (\ref{Sec:MinimalSyst}). The Amount of {\em consistent} information transmitted from $\mathbf{A}$ to $\mathbf{B} -$when $\mathbf{A}$ acts as the coder agent and $\mathbf{B}$ as the decoder one- satisfies that 
\[
{\cal I}(\mathbf{AB})=I(X_{\Omega}:X'_{\Omega}) 
\]
only in the following two extreme cases:
\begin{enumerate}
\item
$I(X_{\Omega}:X'_{\Omega})=0$, or
\item
$\Lambda(\mathbf{A}\mathbf{B})=\delta_{n\times n}$.
\end{enumerate}
Otherwise,
\[
{\cal I}(\mathbf{AB})<I(X_{\Omega}:X'_{\Omega}).
\]
$\\$

{\em Proof}.-The first case is the trivial one in which there is no information available due to total uncertainty -corresponding to  $\epsilon=\frac{1}{2}$ in the case of the symmetric binary channel studied in section \ref{sec:BSC}, see also figure (\ref{Fig:Informations}). The second one is more interesting. Indeed, having $\Lambda(\mathbf{A}\mathbf{B})=\delta$ means that
\[
(\P^{\mathbf{A}},\Lambda,\Q^{\mathbf{B}}\in \Pi_{n\times n})\;{\rm and}\;\P^{\mathbf{A}}=(\Lambda\Q^\mathbf{B})^T,
\]
where we use that, if $\mathbf{C}\in \Pi_{n\times n}$, $\mathbf{C}^{-1}=\mathbf{C}^T$, also having that $\mathbf{C}^T\in\Pi_{n\times n}$. Out of these two situations,  $\exists J_{ik}(\mathbf{AB})>0$, in which $i\neq k$, since there are more than $n$ non-zero entries in the matrix $\Lambda(\mathbf{A}\mathbf{B})$, leading to 
\[
{\cal I}(\mathbf{AB})<I(X_{\Omega}:X'_{\Omega}).
\]
 $\\$

{\em Lemma 2}.-
Let $\mathbf{AB}$ be two agents sharing the world $\Omega$ as described in section (\ref{Sec:MinimalSyst}). The Amount of {\em consistent} information transmitted from $\mathbf{A}$ to $\mathbf{B} -$when $\mathbf{A}$ acts as the coder agent and $\mathbf{B}$ as the decoder one- is bounded as follows:
\begin{equation}
{\cal I}(\mathbf{AB})\leq\left( 1-\frac{H(X_{\Omega}|X'_{\Omega})-1}{\log(n-1)}\right)
\left(\max_i \{D(\Lambda_i (\mathbf{AB})||p')\}\right).
\label{eq:InequalityGen}
\end{equation}
$\\$

{\em Proof}.-Let $\vec{v}$ and $\vec{u}$ be two vectors of $\mathbb{R}^n$. Its scalar product,
\[
\langle \vec{v},\vec{u}\rangle=\sum_{i\leq n}v_iu_i
\]
is bounded, thanks to the so-called {\em H\"older's inequality}, in the following way:
\[
|\langle \vec{v},\vec{u}\rangle|\leq \left(\sum_{i\leq n} v_i^{\alpha}\right)^{\frac{1}{\alpha}}\left(\sum_{i\leq n} u_i^{\beta}\right)^{\frac{1}{\beta}},
\]
as long as $\alpha$ and $\beta$ are {\em H\"older conjugates}, i.e, $1/\alpha+1/\beta=1$. The above expression can be rewritten, using the notation of norms as $|\langle \vec{v},\vec{u}\rangle|\leq ||\vec{v}||_{\alpha}\cdot||\vec{u}||_{\beta}$ -recall that, for $\alpha=\beta=1/2$ we recover the well-known {\em Schwartz inequality} for the euclidean distance. If we put $\alpha \to 1$ and $\beta\to \infty$ we obtain 
\[
|\langle \vec{v},\vec{u}\rangle|\leq ||\vec{v}||_1\cdot||\vec{u}||_{\infty},
\]
where 
\[
||\vec{v}||_1=\sum_{i\leq n} v_i; \;\;{\rm and}\;\;||\vec{u}||_{\infty}=\max_i\{u_i\},
\]
being the last one the so-called {\em Chebyshev's norm}. Now we want to apply this machinery to our problem. The key point is to realize that ${\cal I}(\mathbf{AB})$ can be expressed as a scalar product between two vectors, having the first one coordinates $J_{11}(\mathbf{AB}),...,J_{nn}(\mathbf{AB})$ and the second one $D(\Lambda_1(\mathbf{AB})||q),...,D(\Lambda_n(\mathbf{AB})||p')$. We remark that this step is legitimated because all the terms involved in the computation are positive. Therefore, by applying the H\"older's inequality over the definition of ${\cal I}(\mathbf{AB})$, we have that
\begin{eqnarray}
{\cal I}(\mathbf{AB})&=&\sum_{i\leq n}J_{ii}(\mathbf{AB}) D(\Lambda_i(\mathbf{AB})||p')\nonumber\\
&\leq&\left(\sum_{i\leq n}J_{ii}\right)\left(\max_i \{D(\Lambda_i (\mathbf{AB})||p')\}\right)\nonumber\\
&=&\theta_{\mathbf{AB}}\left(\max_i \{D(\Lambda_i (\mathbf{AB})||p')\}\right),\nonumber
\end{eqnarray} 
being $\theta_{\mathbf{AB}}$ defined in equation (\ref{Eq:Theta}).
Now we observe that the probability of error in referentiating a given event of $\Omega$ is $p_e=1-\theta_{\mathbf{AB}}$. This enables us to use Fano's inequality to bound $\theta_{\mathbf{AB}}$:
\begin{eqnarray}
\theta_{\mathbf{AB}}\leq \left( 1-\frac{H(X_{\Omega}|X'_{\Omega})-1}{\log(n-1)}\right),\nonumber
\end{eqnarray}
thereby obtaining the desired result.
$\\$

{\em Lemma 3}.- (Derivation of inequality (\ref{eq:Inequality2})). Let $\mathbf{AB}$ be two agents sharing the world $\Omega$ as described in section (\ref{Sec:MinimalSyst}) and such that $(\forall m_i\in\Omega)p(m_i)=1/n$ and that the channel defined by $\Lambda(\mathbf{AB})$ is symmetric. Then, the following inequality holds:
\begin{eqnarray}
{\cal I}(\mathbf{AB})<\frac{I^2(X_{\Omega}:X'_{\Omega})}{H(X_{\Omega})}+1.\nonumber
\end{eqnarray}
$\\$

{\em Proof}.-
The first issue is to show that, if $(\forall m_i\in\Omega)p(m_i)=1/n$ and the channel defined by $\Lambda(\mathbf{AB})$ is symmetric, then
$(\forall m_i\in\Omega)$ $ D(\Lambda_i(\mathbf{AB})||q)=I(X_{\Omega}:X'_{\Omega})$. Indeed, since the channel is symmetric $p=p'$ and thus $H(X_{\Omega})=H(X'_{\Omega})=\log n$. Then take any $m_i\in \Omega$ and compute $ D(\Lambda_i(\mathbf{AB})||p')$:
\begin{eqnarray}
D(\Lambda_i(\mathbf{AB})||q)&=&\sum_{j\leq n}\Lambda_{ij}(\mathbf{AB})\log\Lambda_{ij}(\mathbf{AB})+\log n\nonumber\\
&=& \log n -H(X'_{\Omega}|X_{\Omega}=m_i)\nonumber\\
&=&\log n-\sum_{i\leq n}\frac{1}{n}\sum_{j\leq n}H(X'_{\Omega}|X_{\Omega}=m_i)\nonumber\\
&=&I(X_{\Omega}:X'_{\Omega}),\nonumber
\end{eqnarray}
where in the third step we used the property that, in a symmetric channel,  $(\forall m_i, m_j\in\Omega)$ $H(X'_{\Omega}|X_{\Omega}=m_i)=H(X'_{\Omega}|X_{\Omega}=m_j)$. Thus, if we average a constant value, we obtain such a value as the outcome (last step). Then, we apply inequality (\ref{eq:InequalityGen}):
\begin{eqnarray}
{\cal I}(\mathbf{AB})
&\leq&\left( 1-\frac{H(X_{\Omega}|X'_{\Omega})-1}{\log(n-1)}\right)I(X_{\Omega}:X'_{\Omega})\nonumber\\
&<& \left(1- \frac{H(X_{\Omega}|X'_{\Omega})-1}{H(X_{\Omega})}\right)I(X_{\Omega}:X'_{\Omega})\nonumber\\
&\leq&\frac{I^2(X_{\Omega}:X'_{\Omega})}{H(X_{\Omega})}+1,\nonumber
\end{eqnarray}
where, in the second step we used the fact that $H(X_{\Omega})=\log n>\log(n-1)$ and in the third step we bound the remaining term
\[
\frac{I(X_{\Omega}:X'_{\Omega})}{H(X_{\Omega})}\leq 1,
\]
since $I(X_{\Omega}:X'_{\Omega})\leq H(X_{\Omega})$, thus completing the proof.
$\\$

\section{Achieving self-consistency maximizing consistent information}

\label{sec:SelfConsistent}
The structure of the functional accounting for the amount of consistent information shared by two agents -equation (\ref{eq:Symmetrized})- can lead to the paradoxical situation in which high scores on ${\cal I}(\mathbf{A:B})$ do not imply high values on ${\cal I}(\mathbf{A:A})$ or ${\cal I}(\mathbf{B:B})$. In brief, the degeneracy of possible optimal configurations seems to jeopardize self-understanding  even in the case in which communication is optimal. Interestingly, this apparent paradox can be ruled out at the level of populations  of agents, for several representative cases, as demonstrated in \cite{Corominas-Murtra:2006} using a version of $\theta_{\mathbf{AB}}$. For the particular case where $\eta_{AB}=0$, we have seen at the beginning of this section that ${\cal I}(\mathbf{AB})\leq I(X_{\Omega}:X'_{\Omega})$, having equality only in the special case by which $\Lambda({\mathbf{AB}})=\delta_{n\times n}$, which, in turn, implies that  ${\cal I}(\mathbf{AB})=H(X_{\Omega})$. The interesting issue is that in presence of three or more agents $\mathbf{A},\mathbf{B}$ and $\mathbf{C}$:
\[\left.
\begin{array}{lll}
{\cal I}(\mathbf{A}:\mathbf{B})&=&H(X_{\Omega})\\
{\cal I}(\mathbf{A}:\mathbf{C})&=&H(X_{\Omega})\\
{\cal I}(\mathbf{B}:\mathbf{C})&=&H(X_{\Omega})\\
\end{array}\right\}\Rightarrow 
\begin{array}{lll}
{\cal I}(\mathbf{A}:\mathbf{A})&=&H(X_{\Omega})\\
{\cal I}(\mathbf{B}:\mathbf{B})&=&H(X_{\Omega})\\
{\cal I}(\mathbf{C}:\mathbf{C})&=&H(X_{\Omega})\\
\end{array}
\]
i.e., maximizing the communicative success over a  population of agents results automatically in a population of self-consistent agents, although there is no a-priori correlation between the coder and the decoder module of a given agent.
Now we rigorously demonstrate this statement. 
$\\$

{\em Lemma 3}.- Let us have three $\mathbf{A}_i, \mathbf{A}_j, \mathbf{A}_k$ agents communicatively interacting and sharing the world $\Omega$ as described in section (\ref{Sec:MinimalSyst}). Then, if 
\[
(\forall i<k){\cal I}(\mathbf{A}_i:\mathbf{A}_k)=H(X_{\Omega}),
\]
then
\[
(\forall i){\cal I}(\mathbf{A}_i:\mathbf{A}_i)=H(X_{\Omega}).
\]
$\\$

{\em Proof}.-
We observe, as discussed above, that the premise only holds if $(\forall i<k)$
\[
\P^{\mathbf{A}_i},\Q^{\mathbf{A}_i}, \Lambda, \P^{\mathbf{A}_k}, \Q^{\mathbf{A}_k}\in \Pi_{n\times n},
\]
and 
\[
\P^{\mathbf{A}_i}=(\Lambda \Q^{\mathbf{A}_k})^T\wedge \P^{\mathbf{A}_k}=(\Lambda \Q^{\mathbf{A}_i})^T.
\]
Now we observe that, if ${\cal I}(\mathbf{A}_i:\mathbf{A}_k)=H(X_{\Omega})$, ${\cal I}(\mathbf{A}_i:\mathbf{A}_j)=H(X_{\Omega})$, we conclude that:
\[
\Q^{\mathbf{A}_k}=\Q^{\mathbf{A}_j}; \wedge \P^{\mathbf{A}_k}=\P^{\mathbf{A}_j}
\]
i.e., $\mathbf{A}_k=\mathbf{A}_j$. Now, knowing that 
${\cal I}(\mathbf{A}_k:\mathbf{A}_j)=H(X_{\Omega})$, then:
\[
{\cal I}(\mathbf{A}_k:\mathbf{A}_k)=H(X_{\Omega}).
\]
We can easily generalize such a reasoning over an arbitrarily large number of communicating agents.  


\begin{thebibliography}{19}
\expandafter\ifx\csname natexlab\endcsname\relax\def\natexlab#1{#1}\fi
\expandafter\ifx\csname bibnamefont\endcsname\relax
  \def\bibnamefont#1{#1}\fi
\expandafter\ifx\csname bibfnamefont\endcsname\relax
  \def\bibfnamefont#1{#1}\fi
\expandafter\ifx\csname citenamefont\endcsname\relax
  \def\citenamefont#1{#1}\fi
\expandafter\ifx\csname url\endcsname\relax
  \def\url#1{\texttt{#1}}\fi
\expandafter\ifx\csname urlprefix\endcsname\relax\def\urlprefix{URL }\fi
\providecommand{\bibinfo}[2]{#2}
\providecommand{\eprint}[2][]{\url{#2}}


\bibitem[{\citenamefont{Schuster}(2001)}]{Schuster:2001}
\bibinfo{author}{\bibnamefont{Schuster}, \bibfnamefont{P.}},
  \bibinfo{year}{2001}, \bibinfo{journal}{Front. Life}
  \textbf{\bibinfo{volume}{}}(\bibinfo{number}{1}), \bibinfo{pages}{329-346
  
\bibitem{Szathmary:1997}
Maynard Smith, J,  Sz\'athm\'ary, E., 1997,
{\em The major transitions in evolution}
(Oxford U. Press, Oxford.)

\bibitem{MaynardSmith:2000}
Maynard Smith, J. 2000,
Philosophy of science (67), 177-194

\bibitem[{\citenamefont{Hopfield}(1994)}]{Hopfield:1994}
\bibinfo{author}{\bibnamefont{Hopfield}, \bibfnamefont{J.}},
  \bibinfo{year}{1994}, \bibinfo{journal}{Journal of Theoretical Biology}
  \textbf{\bibinfo{volume}{171}}(\bibinfo{number}{1}), \bibinfo{pages}{53}

\bibitem[{\citenamefont{Shannon}(1948)}]{Shannon:1948}
\bibinfo{author}{\bibnamefont{Shannon}, \bibfnamefont{C.~E.}},
  \bibinfo{year}{1948}, \bibinfo{journal}{Bell System Technical Journal}
  \textbf{\bibinfo{volume}{27}}, \bibinfo{pages}{379}.
  
  \bibitem{Bialek:2012}
Bialek, W., 2012,
{\em Biophysics: searching for principles}, 
(Princeton University Press.)

\bibitem[{\citenamefont{Saussure}(1916)}]{Saussure:1916}
\bibinfo{author}{\bibnamefont{Saussure}, \bibfnamefont{F.}},
  \bibinfo{year}{1916}, \emph{\bibinfo{title}{Cours de Linguistique
  G\'en\'erale}} (\bibinfo{publisher}{Biblioth\`eque scientifique Payot:
  Paris})}

\bibitem{Atlan:1987}
Atlan, H., 1987, Physica Scripta (36), 563-576.

\bibitem{Atlan:1998}
Atlan, H., 1998,
International Immunology,  (6), 711-717.

\bibitem[{\citenamefont{Hurford}(1989)}]{Hurford:1989}
\bibinfo{author}{\bibnamefont{Hurford}, \bibfnamefont{J.}},
  \bibinfo{year}{1989}, \bibinfo{journal}{Lingua}
  \textbf{\bibinfo{volume}{77}}(\bibinfo{number}{2}), \bibinfo{pages}{187}

\bibitem[{\citenamefont{Komarova and Niyogi}(2004)}]{Komarova:2004}
\bibinfo{author}{\bibnamefont{Komarova}, \bibfnamefont{N.~L.}}, and
  \bibinfo{author}{\bibfnamefont{P.}~\bibnamefont{Niyogi}},
  \bibinfo{year}{2004}, \bibinfo{journal}{Art. Int.}
  \textbf{\bibinfo{volume}{154}}(\bibinfo{number}{1-2}), \bibinfo{pages}{1},

\bibitem[{\citenamefont{Niyogi}(2006)}]{Niyogi:2006}
\bibinfo{author}{\bibnamefont{Niyogi}, \bibfnamefont{P.}},
  \bibinfo{year}{2006}, \emph{\bibinfo{title}{The Computational Nature of
  Language Learning and Evolution}} (\bibinfo{publisher}{MIT Press. Cambridge,
  Mass.}),

\bibitem[{\citenamefont{Nowak}(2000)}]{Nowak:2000}
\bibinfo{author}{\bibnamefont{Nowak}, \bibfnamefont{M.~A.}},
  \bibinfo{year}{2000}, \bibinfo{journal}{Philosophical Transactions:
  Biological Sciences} \textbf{\bibinfo{volume}{355}}(\bibinfo{number}{1403}),
  \bibinfo{pages}{1615},

\bibitem[{\citenamefont{Nowak and Krakauer}(1999)}]{Nowak:1999}
\bibinfo{author}{\bibnamefont{Nowak}, \bibfnamefont{M.~A.}}, and
  \bibinfo{author}{\bibfnamefont{D.}~\bibnamefont{Krakauer}},
  \bibinfo{year}{1999}, \bibinfo{journal}{Proc. Nat. Acad. Sci. USA}
  \textbf{\bibinfo{volume}{96}}(\bibinfo{number}{14}), \bibinfo{pages}{8028},

\bibitem[{\citenamefont{Plotkin and Nowak}(2000)}]{Plotkin:2000}
\bibinfo{author}{\bibnamefont{Plotkin}, \bibfnamefont{J.~B.}}, and
  \bibinfo{author}{\bibfnamefont{M.~A.} \bibnamefont{Nowak}},
  \bibinfo{year}{2000}, \bibinfo{journal}{Journal of Theoretical Biology}
  \textbf{\bibinfo{volume}{205}}(\bibinfo{number}{1}), \bibinfo{pages}{147},

\bibitem{Cangelosi:2002}
Cangelosi A., Parisi D., (Eds.) , 2002, {\em Simulating the Evolution of Language} Springer: London.

\bibitem{Floreano:2007}
Floreano, D., Mitri, S., Magnenat, S., Keller, L., 2007,  Curr.
Biol., (17), 514-519.

\bibitem[{\citenamefont{Steels}(2001)}]{Steels:2001}
\bibinfo{author}{\bibnamefont{Steels}, \bibfnamefont{L.}},
  \bibinfo{year}{2001}, \bibinfo{journal}{IEEE Intelligent Systems}
  \textbf{\bibinfo{volume}{16}}, \bibinfo{pages}{16}, ISSN
  \bibinfo{issn}{1541-1672}.

\bibitem[{\citenamefont{Steels and Baillie}(2003)}]{Steels:2003}
\bibinfo{author}{\bibnamefont{Steels}, \bibfnamefont{L.}}, and
  \bibinfo{author}{\bibfnamefont{J.-C.} \bibnamefont{Baillie}},
  \bibinfo{year}{2003}, \bibinfo{journal}{Robotics and Autonomous Systems}
  \textbf{\bibinfo{volume}{43}}(\bibinfo{number}{2-3}), \bibinfo{pages}{163}.
 
  \bibitem{Steels:2003b}
Steels, L., 2003,
Trends. Cogn. Sci., (7), 308-312.

\bibitem{Steels:2005}
Steels, L., 2005, Connect. Sci., (17), 213-230.

\bibitem[{\citenamefont{Nolfi and Mirolli}(2010)}]{Nolfi:2010}
\bibinfo{author}{\bibnamefont{Nolfi}, \bibfnamefont{S.}}, and
  \bibinfo{author}{\bibfnamefont{M.}~\bibnamefont{Mirolli}},
  \bibinfo{year}{2010}, \emph{\bibinfo{title}{Evolution of Communication and
  Language in Embodied Agents}} (\bibinfo{publisher}{Berlin. Springer Verlag})

\bibitem[{\citenamefont{Ash}(1990)}]{Ash:1990}
\bibinfo{author}{\bibnamefont{Ash}, \bibfnamefont{R.~B.}},
  \bibinfo{year}{1990}, \emph{\bibinfo{title}{Information Theory}}
  (\bibinfo{publisher}{New York. Dover}).

\bibitem[{\citenamefont{Cover and Thomas}(1991)}]{Thomas:2001}
\bibinfo{author}{\bibnamefont{Cover}, \bibfnamefont{T.~M.}}, and
  \bibinfo{author}{\bibfnamefont{J.~A.} \bibnamefont{Thomas}},
  \bibinfo{year}{1991}, \emph{\bibinfo{title}{Elements of Information Theory}}
  (\bibinfo{publisher}{John Wiley and Sons. New York})
   
\bibitem{Gleick:2012}
Gleick, J. 2012
{\em The information: A history, a theory, a flood. },
(Vintage, New York.)
    
\bibitem{Bickerton:1990}
Bickerton, D., 1990 {\em Language and Species}, (Chicago University Press: Chicago.)

\bibitem{Christiansen:2003}
Christiansen, M.H., Kirby, S., 2003  Trends. Cogn. Sci., (7), 300-307.  

\bibitem{Deacon:1997}
Deacon, T.W., 1997, {\em The Symbolic Species: The Co-Evolution of Language and the Brain} (Norton: New York.)
  
\bibitem{Corominas-Murtra:2006}
Corominas-Murtra, B. and Sol\'e, R. V. 2006,  Journal of Theoretical Biology, 241(2):438--441.





  


\bibitem[{\citenamefont{Haken}(1978)}]{Haken:1978}
\bibinfo{author}{\bibnamefont{Haken}, \bibfnamefont{H.}}, \bibinfo{year}{1978},
  \emph{\bibinfo{title}{Synergetics: An Introduction. Nonequilibrium Phase
  Transitions and Self- Organization in Physics, Chemistry and Biology
  (Springer Series in Synergetics)}} (\bibinfo{publisher}{Springer}). 



\bibitem[{\citenamefont{Kolmogorov}(1965)}]{Kolmogorov:1965}
\bibinfo{author}{\bibnamefont{Kolmogorov}, \bibfnamefont{A.}},
  \bibinfo{year}{1965}, \bibinfo{journal}{Problems Inform. Transmission}
  \textbf{\bibinfo{volume}{1}}, \bibinfo{pages}{1}.


\bibitem[{\citenamefont{Ming and Vit\'{a}nyi}(1997)}]{Li:1997}
\bibinfo{author}{\bibnamefont{Ming}, \bibfnamefont{L.}}, and
  \bibinfo{author}{\bibfnamefont{P.}~\bibnamefont{Vit\'{a}nyi}},
  \bibinfo{year}{1997}, \emph{\bibinfo{title}{An introduction to Kolmogorov
  complexity and its applications}} (\bibinfo{publisher}{Springer},
  \bibinfo{address}{New York [u.a.]}),



\end{thebibliography}

\end{document}